\documentclass[aps,pre,twocolumn,showpacs,showkeys,amsfonts,amssymb,amsmath,letterpaper]{revtex4}
\usepackage[dvips]{graphicx}

\begin{document}

\title{Supersymmetric Langevin equation to explore free energy landscapes}
\author{Alessandro Mossa}
\author{Cecilia Clementi}
\affiliation{Department of Chemistry, Rice University, 6100 Main Street, Houston, Texas 77005}
\date\today

\begin{abstract}
The recently discovered supersymmetric generalizations of Langevin dynamics and Kramers equation 
can be utilized for the exploration of free energy landscapes of systems whose large time-scale 
separation hampers the usefulness of standard molecular dynamics techniques.   
The first realistic application is here presented. The system chosen is a minimalist model for a 
short alanine peptide exhibiting a helix-coil transition. 
\end{abstract}

\pacs{02.70.Ns, 87.15.Aa, 05.10.Gg}
% 87.15.Aa: Biomolecules: structure and physical properties - Theory and modeling; computer simulation
% 02.70.Ns: Molecular dynamics and particle methods
% 05.10.Gg: Stochastic analysis methods (Fokker-Planck, Langevin, etc.)
\keywords{Reaction paths, Molecular dynamics, Supersymmetry, Helix-coil transition, Time-scale separation}

\maketitle

The problem of identifying and exploring critical regions on free energy 
landscapes is central to many disciplines lying at the interface between
physics and chemistry \cite{Wales2003}: It is an outstanding issue in the study 
of atomic clusters, glasses, supercooled liquids, biopolymer dynamics, 
and protein folding. 

Many authors \cite{JCP85_5045,PRL78_3908,JPCB103_899,PRL87_108302,ARPC53_291,PRB66_052301,
PRL97_108101} 
have been developing methods to find reaction paths 
based on a statistical description of the ensemble of pathways connecting certain 
phase space regions. A radically different approach has been recently proposed 
\cite{PRL91_188302,JSP116_1201,JSP122_557}: The supersymmetry hidden 
in the Kramers equation can be made explicit by coupling to the ``bosonic'' 
degrees of freedom in the phase space an equal number of
``fermionic'' ones in the tangent space. As the usual Kramers equation is the 
base of molecular dynamics (henceforth simply MD) simulations, so its 
supersymmetric generalization gives rise to an enhanced MD (let's call it 
SuSy MD), suitable for the study of systems characterized by time-scale 
separation. Rigorous theoretical arguments and 1- and 2-dimensional toy model 
applications have established that, in a purely energetic landscape, SuSy MD 
is able to find reaction paths in a time much shorter than the activation 
time.
 
However, in order to pave the way to realistic problem applications, the 
feasibility of SuSy MD must be established in the framework of high dimensional 
\emph{free energy} landscapes. By studying a model not very computationally 
demanding, but possessed of all the other complications which characterize current 
research in biomolecular simulations, 
we provide the missing piece of evidence that the new method is able to signal the 
presence of entropic as well as energetic barriers. We show that the identification 
of the reaction path obtained by SuSy MD is fully consistent with the results of 
standard MD, with the advantage that the simulation time needed is orders of 
magnitude shorter.
 
In the following we first briefly review the supersymmetric formulation of 
Kramers equation, urging the reader interested in more details to peruse 
Refs.~\cite{JSP116_1201,JSP122_557}: With respect to the existing 
literature, however, we try to clarify the peculiar problems raised by 
the application to free energy landscapes. A brief description of technical 
aspects of the method used is followed by the results of its application to 
our test system.
    
\section{Supersymmetric Kramers equation}

We consider a system of $n$ interacting particles in the 3-dimensional space, 
defined by the Hamiltonian
\begin{equation}
  \mathcal{H}=\frac{\mathbf{p}^2}{2m}+V(\mathbf{q}) \,,
\end{equation}
where the vectors $\mathbf{q} = (\vec{q}_{1}, \dots, \vec{q}_{n})$ and $\mathbf{p} =
(\vec{p}_{1}, \dots, \vec{p}_{n})$ indicate the positions and momenta associated to the 
particles, and $V(\mathbf{q})$ is the interaction potential. 
To simplify the notation, we assign to each particle the same mass $m$. 
The dynamics of the system coupled to a heat bath at constant temperature $T$  
is described by means of a Langevin equation
\begin{equation} \label{Langevin}
  \left\{
  \begin{array}{rcl}
     \dot{\mathbf{q}} &=& \mathbf{p}/m \\
     \dot{\mathbf{p}} &=& -\nabla V+\sqrt{2m\gamma T}\boldsymbol{\eta}-\gamma \mathbf{p} \,,
  \end{array}  
  \right.
\end{equation}
where we have fixed the Boltzmann constant $k_\mathrm{B}=1$, the friction coefficient is $\gamma$,
and $\boldsymbol{\eta}$ is a Gaussian white noise:
\begin{eqnarray}
  \langle\eta_\mu(t)\rangle&=&0 \\
  \langle\eta_\mu(t)\eta_\nu(t')\rangle&=&\delta_{\mu\nu}\delta(t-t') \,.
\end{eqnarray}
The indices $\mu$ and $\nu$ run over all the configuration space degrees of freedom $1,\dots,N$, with $N=3n$.

The phase space probability density $W({\bf q},{\bf p},t)$ evolves according 
to the Kramers equation \cite{Risken1996} 
\begin{equation} \label{Kramers}
  \frac{\partial}{\partial t}W({\bf q},{\bf p},t)=-H_\mathrm{K}W({\bf q},{\bf p},t) \,,
\end{equation}
where
\begin{equation} \label{H_K}
  H_\mathrm{K}=\sum_{\mu=1}^{N}\left[\frac{\partial}{\partial
  q_\mu}\frac{p_\mu}{m}-\frac{\partial}{\partial p_\mu}\left(m\gamma
  T\frac{\partial}{\partial p_\mu}+\gamma p_\mu+\frac{\partial V}{\partial
  q_\mu}\right)\right] \,.
\end{equation}
The Kramers equation can be rewritten as a continuity equation for the
probability current \cite{Risken1996}
\begin{eqnarray} \label{prob_curr}
  J_{q_\mu}&=&\frac{p_\mu}{m}W(\mathbf{q},\mathbf{p},t) \\
  J_{p_\mu}&=&-\left(m\gamma T\frac{\partial}{\partial p_\mu}+\gamma
  p_\mu+\frac{\partial V}{\partial q_\mu}\right) W(\mathbf{q},\mathbf{p},t) \,.
  \nonumber 
\end{eqnarray}
It has been shown \cite{PLA235_105,JSP122_557} that 
a hidden supersymmetry is associated with the Kramers equation: 
By extending the space with $4N$ fermion operators
\begin{equation}
  \{a_\mu,a_\nu^\dag\}=\delta_{\mu\nu} \qquad \qquad \qquad \{b_\mu,b_\nu^\dag\}=\delta_{\mu\nu} \,,
\end{equation}
a supersymmetric extension of Eq.~(\ref{Kramers}) is obtained 
\begin{equation} \label{H}
  H_\mathrm{SK}=H_\mathrm{K}+\frac{1}{m}\sum_{\mu,\nu =1}^{N}\frac{\partial^2 V}{\partial
  q_\mu\partial q_\nu}b_\mu^\dag a_\nu+ \sum_{\mu=1}^{N}\left( \gamma b_\mu^\dag b_\mu-a_\mu^\dag b_\mu \right)
  \,.
\end{equation}
By defining a $2N$-component vector ${\bf x}$ such that
$x_\mu=q_\mu$ and $x_{N+\mu}=p_\mu$, for $\mu=1,\dots,N$\,
the evolution operator Eq.~(\ref{H}) can be expressed in compact notation
\begin{equation}
  H_\mathrm{SK}=H_\mathrm{K}+\sum_{i,j=1}^{2N}A_{ij}c^\dag_i c_j \label{H_sk} \,,
\end{equation}
where $(c_1,\dots,c_{2N})=(a_1,\dots,a_N,mb_1,\dots,mb_N)$, and the matrix $A$ is
\begin{equation} \label{Adef}
  A=\left(
    \begin{array}{cc}
      0 & -\delta_{\mu\nu}/m \\
      \frac{\partial^2V}{\partial q_\mu\partial q_\nu} & \gamma \delta_{\mu\nu}
    \end{array}
  \right) \,.
\end{equation}
The solution to the supersymmetric version of Eq.~(\ref{Kramers}) can 
be expressed in the form
\begin{equation} \label{wavefunction}
|\psi^{(k)}(\mathbf{x},t)\rangle=\sum_{i_1,\dots,i_k}\psi_{i_1,\dots,i_k}(\mathbf{x},t) \ 
 c_{i_1}^\dag\cdots c_{i_k}^\dag|-\rangle \,,
\end{equation}   
where the function $\psi_{i_1,\dots,i_k}(\mathbf{x},t)$ has the physical meaning of
probability density in the phase space, $|-\rangle$ is the fermion vacuum, 
and $k$ is the fermion number, that is, an eigenvalue of the operator
$N_\mathrm{f}=\sum_i c_i^\dag c_i$. By using this notation,  
the supersymmetric extension of Eq.~(\ref{Kramers}) is written as
\begin{equation} \label{SusyKramers}
  \frac{\partial}{\partial t}|\psi^{(k)}(\mathbf{x},t)\rangle=
  -H_\mathrm{SK}|\psi^{(k)}(\mathbf{x},t)\rangle \,.
\end{equation}

\section{Supersymmetric molecular dynamics}
Let us first consider the solution to Eq.~(\ref{SusyKramers}) in the 
zero-fermion sector, where 
$|\psi^{(0)}(\mathbf{x},t)\rangle=W(\mathbf{x},t)|-\rangle$. In this case
we simply recover the Kramers equation (\ref{Kramers}).
If we start from some initial condition $|\psi(\mathbf{x},0)\rangle$,
we can expand the generic state $|\psi(\mathbf{x},t)\rangle$ into right eigenvectors
$|\psi^\mathrm{R}_\alpha(\mathbf{x})\rangle$ of the operator $H_{\mathrm{K}}$
\begin{equation}
  |\psi(\mathbf{x},t)\rangle=\sum_\alpha C_\alpha(t)|\psi^\mathrm{R}_\alpha(\mathbf{x})\rangle \,,
\end{equation}
so that Eq.~(\ref{Kramers}) yields
\begin{equation}
  |\psi(\mathbf{x},t)\rangle=\sum_\alpha C_\alpha(0)e^{-\lambda_\alpha t}|\psi_\alpha^\mathrm{R}(\mathbf{x})\rangle \,, 
\end{equation}
where $H_\mathrm{K}|\psi_\alpha^\mathrm{R}\rangle=\lambda_\alpha|\psi_\alpha^\mathrm{R}\rangle$.
As $t$ increases, this sum is obviously more and more dominated by the
eigenvectors with the smallest eigenvalues. For $t\to\infty$, only the
stationary state (defined by $\lambda=0$) survives. If the system is
characterized by the presence of two (or more) \emph{well separated time-scales}
$\tau_\mathrm{fast} \ll \tau_\mathrm{slow} $, a corresponding gap 
is also present in the spectrum of $H_\mathrm{K}$.

It follows that at a time $\tilde{t}$ such that
$\tau_\mathrm{fast}\ll\tilde{t}\ll\tau_\mathrm{slow}$, the evolution of the system 
is well approximated by a linear superposition of the $K$ right eigenvectors
below the gap:
\begin{equation}
  |\psi(\mathbf{x},\tilde{t})\rangle\approx\sum_{\alpha=0}^{K-1}
  C_\alpha(0)e^{-\lambda_\alpha
  \tilde{t}}|\psi_\alpha^\mathrm{R}(\mathbf{x})\rangle \,.
\end{equation}

In the framework of the master equation formulation of non-equilibrium
statistical mechanics, it can be proved \cite{JMP37_3897,PRE64_016101,CMP228_219} that $K$ 
suitable linear combinations of the right eigenvectors $|\psi_\alpha^\mathrm{R}\rangle$ 
below the gap exist such that the associated probability densities
$W(\mathbf{q},\mathbf{p},t)$ are positive normalized distributions, nonzero only
on non-overlapping regions of the configuration space, 
and stationary on time-scales much shorter than $\tau_\mathrm{slow}$. 
One can therefore use these states for a rigorous and general 
\emph{definition} of metastability.  It is important to stress that this
results hold true independently on the origin of the time-scale separation.

The probability distribution $W(\mathbf{x},t)$ can be used to define a 
\textit{dynamic free energy}:
\begin{equation} \label{freeenergy}
  \mathcal{F}(t)=\int \frac{\mathrm{d}^{2N}\mathbf{x}}{h^N}\left[
  \mathcal{H}(\mathbf{x})W(\mathbf{x},t)+TW(\mathbf{x},t)\ln W(\mathbf{x},t)\right]\, .
\end{equation}
For $t\to\infty$ the probability distribution tends to the Boltzmann distribution
\begin{equation}
 \lim_{t\to\infty}W(\mathbf{q},\mathbf{p},t)=
 \frac{1}{Z}\exp\left(-\frac{\mathcal{H}(\mathbf{q},\mathbf{p})}{T}\right) \,,
\end{equation}
where $Z$ is the partition function. It follows that
\begin{equation}
  \lim_{t\to\infty}\mathcal{F}(t)=-T\ln Z \,,
\end{equation}
that is the equilibrium definition of the Helmholtz free energy in the canonical ensemble.
Usual constant temperature MD simulations are limited to the study of the 
zero-fermion sector: for a given potential 
$V(\mathbf{q})$ the Langevin equation is numerically integrated for a time 
$\tilde{t} \gg \tau_\mathrm{slow}$ large enough to reach equilibrium. 

In the one-fermion sector, on the other hand, the wavefunction
Eq.~(\ref{wavefunction}) reads 
\begin{equation}
  |\psi^{(1)}(\mathbf{x},t)\rangle=\sum_{i=1}^{2N}\psi_i(\mathbf{x},t)\  c_i^\dag\ |-\rangle \,,
\end{equation} 
and Eq.~(\ref{SusyKramers}) may be written as
\begin{equation} \label{SK.1ferm}
  \frac{\partial}{\partial t}\psi_i(\mathbf{x},t)=-H_\mathrm{K}\psi_i(\mathbf{x},t)-\sum_{j=1}^{2N}A_{ij}\psi_j(\mathbf{x},t) 
\end{equation}
where we have made use of Eq.~(\ref{H_sk}).
This equation can be solved with the ansatz 
$\psi_i(\mathbf{x},t)=\varphi(\mathbf{x},t)w_i(t)$, where
$\mathbf{w}$ is a vector of dimension $2N$ that does not depend on
$\mathbf{x}\equiv(\mathbf{q},\mathbf{p})$, and $\varphi(\mathbf{x},t)$ 
evolves with the Kramers equation 
\begin{equation}
  \frac{\partial}{\partial
  t}\varphi(\mathbf{x},t)=-H_\mathrm{K}\varphi(\mathbf{x},t) \,.
\end{equation}
This leaves for the vector $\mathbf{w}$ the evolution equation
\begin{equation}\label{eq:dwdt}
  \frac{\mathrm{d}}{\mathrm{d} t}w_i=-\sum_{j=1}^{2N}A_{ij}w_j \,.
\end{equation}
In order to avoid a divergence of the norm 
of $\mathbf{w}$, Eq.~(\ref{eq:dwdt}) can be modified by adding a term: 
\begin{equation} \label{eqcompass}
  \frac{\mathrm{d}}{\mathrm{d}t}w_i=\mathcal{N}(\mathbf{w})w_i-\sum_{j=1}^{2N}A_{ij}w_j \,.
\end{equation}
The norm $|\mathbf{w}|$ is now constant provided that we choose
\begin{equation} \label{rate}
 \mathcal{N}(\mathbf{w})=
 \frac{\mathbf{w}^\mathrm{t}A\mathbf{w}}{|\mathbf{w}|^2} \,.
\end{equation}

The joint distribution $W(\mathbf{x},\mathbf{w},t)$ evolves according to
\begin{equation}
\begin{split}
\frac{\partial W}{\partial t}  & = \left[-H_\mathrm{K} - \mathcal{N}(\mathbf{w}) + \right. \\
 & + \left. \sum_{i=1}^{2N} \frac{\partial}{\partial w_i} \left( \sum_{j=1}^{2N}
  A_{ij}w_j-\mathcal{N}(\mathbf{w})w_i\right)\right]W \,, \\
\label{joint}
\end{split}
\end{equation}
as can be checked by defining
\begin{equation}
  \psi_i(\mathbf{x},t)=\int \mathrm{d}^{2N}\mathbf{w}\,w_i W(\mathbf{x},\mathbf{w},t) 
\end{equation}
and integrating by parts. 

The rules of SuSy MD are easily
read from the RHS of Eq.~(\ref{joint}). We are going to explore the free energy
landscape by means of ``walkers'' moving around in the phase space according to
the usual Langevin dynamics (first term) and each walker carries a ``compass''
$\mathbf{w}$ which evolves with Eq.~(\ref{eqcompass}) (third term). The second
term tells us that the number of walkers grows or decreases with rate
$-\mathcal{N}(\mathbf{w})$. 

How does the presence of different time-scales reflect in the 1-fermion sector 
of the spectrum of $H_\mathrm{SK}$? 
In a simplified setting where entropy plays no role and the separation of
time-scales is purely due to the characteristics of the energy landscape, the
use of a WKB technique in the limit $T\to 0$ shows explicitly
\cite{JSP116_1201} that, while the 0-fermion states are Gaussians centered on
the local minima of the energy, the correspondent (i.e. related by
the supersymmetry) 1-fermion states are the ``reduced current'' densities
\cite{JSP122_557} (obtained by applying the SuSy charge operator to the
probability currents (\ref{prob_curr})), concentrated on the saddles that
separate those minima. In other words, the dynamics given by
Eqs.~(\ref{Langevin}, \ref{eqcompass}) evolves in such a way that the walkers
quickly (that is, on a time-scale larger than $\tau_\mathrm{fast}$ but much
smaller than $\tau_\mathrm{slow}$) organize themselves into trails going from
one local minimum to another one by overcoming the energy barrier along the
reaction path \cite{PRL91_188302}. 

\begin{figure}
\includegraphics[width=\linewidth]{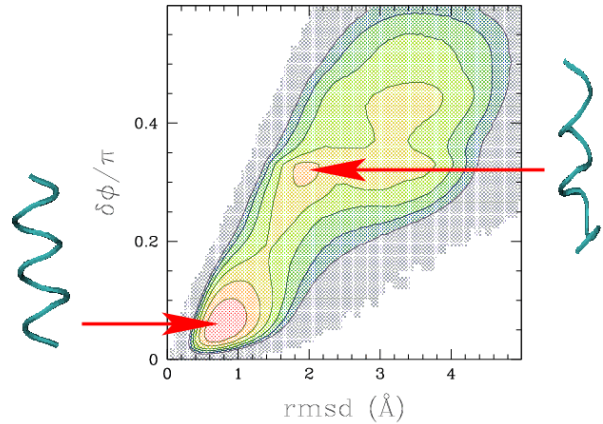}
\caption{\label{fig.helix} The free energy profile at the folding temperature, as 
obtained from equilibrium MD simulations.  Representative structures are shown 
for the helix-shaped native state and the unfolded state. 
Each contour marks an increase of free energy of 1 Kcal/mol.}
\end{figure}
 
Since in the zero-fermion sector the right eigenvectors below the gap 
define the metastable states independently on the physical source of metastability,
it is tempting to speculate that the interpretation of 1-fermion low-lying
states as reaction paths holds also for the general case involving entropy, with a single reaction
path in the free energy landscape standing now for a collection of paths in the
phase space. As a matter of fact, in the zero-temperature limit the dynamic free
energy Eq.~(\ref{freeenergy}) reduces to the energy, therefore we can think of
the WKB argument in Ref.~\cite{JSP122_557} as a rigorous proof, albeit given in a 
limiting case, of a more general statement. While the generalization of the proof 
to finite temperatures is currently in progress, we support here the validity of 
these ideas by showing that indeed SuSy MD can be used to efficiently identify 
reaction paths and saddle points on a free energy landscape, in a system where 
both entropic and energetic factors play a role.

\section{The helix-coil transition}

\begin{figure}
\includegraphics[width=\linewidth]{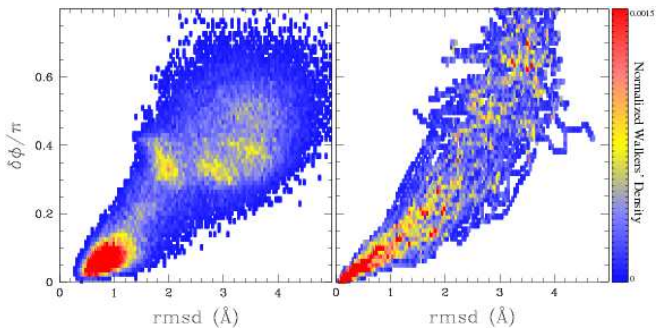}
\caption{
Relative walkers' density at $t=0$ for two very different 
initial conditions, that lead to the same final result. The two initial distributions 
of walkers have been generated by: (Case A, shown in the left panel) Performing
long equilibrium simulations around the transition temperature $T_\mathrm{f}$,
and (Case B, shown in the right panel) performing very short unfolding simulations 
at a much higher temperature $T \gg T_\mathrm{f}$. \label{fig.t0} }
\end{figure}

The choice of the helix-coil transition as test system is a natural one:
It is a simple phenomenon, theoretically well understood \cite{JCP31_526}, 
whose free energy is shaped by the competition between energy and 
entropy into a landscape with two well defined minima, corresponding to the
folded and unfolded states (see Fig.~\ref{fig.helix}). At the transition
temperature $T_\mathrm{f}$ the two minima are equally populated.
By using a coarse-grained off-lattice model we keep relatively low the
dimensionality of the phase space (72 degrees of freedom for our 12-monomer 
chain, see the Appendix for detail), thus reducing the computational
effort while retaining the relevant physical features of a typical two-state folder.

In order to visualize the results, we need to project the 72-dimensional phase
space associated to our model onto a lower dimensional space spanned by a few 
reaction coordinates $\boldsymbol{\xi}$.  
Although the definition of appropriate reaction coordinates for the 
characterization of multidimensional biophysical processes is in general an
area of active research~\cite{PNAS103_9885}, 
a fairly natural set of coordinates is associated to the simple
helix-coil transition considered here: The root mean square deviation (rmsd) 
\cite{ACA32_922,ACA34_827} from the native state $\mathbf{x}^{(0)}$
\begin{equation}
 \mathrm{rmsd}(\mathbf{x},\mathbf{x}^{(0)})=
 \min_{R\in\mathsf{SO}(3)}\textstyle{\frac{1}{2}}|(R\mathbf{x}-\mathbf{x}^{(0)})|^2 \,,
\end{equation}
and the ``helicity'' $\delta \phi$ \cite{PNAS102_14569}
\begin{equation}
  \delta\phi = \sqrt{\sum_{i=1}^{N_\mathrm{R}-3}(\phi_i-\phi_i^{(0)})^2/(N_\mathrm{R}-3)} \,,
\end{equation}  
where $N_\mathrm{R}$ is the number of residues and $\phi_i,\phi_i^{(0)}$ are
the dihedral angles of a generic configuration and of the native configuration,
respectively. 

As the position $\mathbf{x}$ is projected onto the space spanned by the
reaction coordinates $\boldsymbol{\xi}$, so is the vector $\mathbf{w}$, by
means of the Jacobian matrix:
\begin{equation}
  \omega^n= \sum_{i}\frac{\partial \xi^n}{\partial x^i}w^i \,, 
\end{equation}
where $\boldsymbol{\omega}$ is the projected vector. 

The study of the system by means of SuSy MD first requires the
generation of an initial distribution of a large number of walkers in the
accessible phase space.  Each walker $(\mathbf{x},\mathbf{w})$ is then evolved
independently according to Eqs.~(\ref{Langevin}) and (\ref{eqcompass}). 
Moreover, after each time step $\delta t$, there is a probability
$|\mathcal{N}(\mathbf{w})\delta t|$ for every walker of being eliminated if
$\mathcal{N}(\mathbf{w})>0$ or cloned if $\mathcal{N}(\mathbf{w})<0$. 

\begin{figure}
\includegraphics[width=\linewidth]{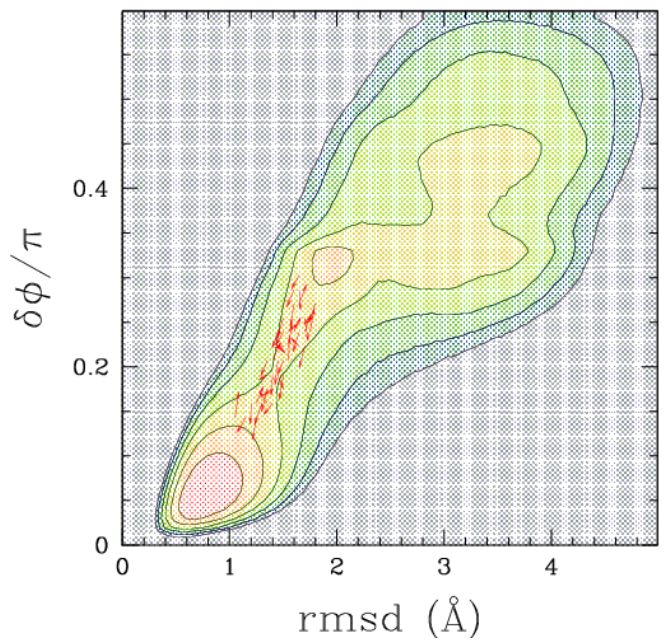}
\caption{The walkers' distribution obtained by SuSy MD (as describd in the
text) identifies the transition state region and reaction paths for the helix-coil
transition. The red arrows illustrate the orientation of the compasses associated 
to the walkers, and are superimposed to the independently determined free 
energy profile, at the transition temperature $T_\mathrm{f}$. Each contour 
marks an increase of free energy of 1 Kcal/mol. A logarithmic scale has been
adopted to improve the readability of the figure: If an arrow in the picture is 
twice longer, the actual norm of the vector is ten times larger.
\label{fig.vecs} }
\end{figure}

Figure~\ref{fig.t0} shows two different distributions of walkers' initial
configurations that have been used in this study. The distributions were
generated by running MD simulations in very different conditions: The
initial distribution of walkers shown in the left panel (case A, in the following)
is obtained by performing equilibrium simulations around the folding 
temperature $T\simeq T_\mathrm{f}$, over a very long timescale $\Delta t
\gtrsim \tau_\mathrm{slow}$, so that the initial walkers' distribution mirrors
faithfully the free energy landscape. On the contrary, the distribution shown
in the right panel (case B) corresponds to configurations sampled during 
very rapid ($\Delta t \lesssim \tau_\mathrm{fast}$) unfolding simulations at a 
temperature $T \gg T_\mathrm{f}$. Our experience is that 
the initial distribution of walkers does not affect the result, as long as the
region of the landscape between the folded and the unfolded state is fairly populated.

In principle, the transition region is simply revealed by the
alignment of the compasses: They are randomly oriented within the states, while
along the transition path they display coherent behavior. In practice, one
needs to sift the points according to criteria such as the walkers density and
the average rate $\mathcal{N}(\mathbf{w})$. After thorough testing, we
have selected an analysis protocol consisting of the following three steps:
\begin{enumerate}
  \item Select a time window;
  \item Select a density threshold;
  \item Select a threshold value for the variance of $\vartheta$ (defined below).
\end{enumerate}
In the following section we detail each step, showing all the phases of the process
which leads from the raw data to the emergence of the reaction path. 

\begin{figure}
\includegraphics[width=\linewidth]{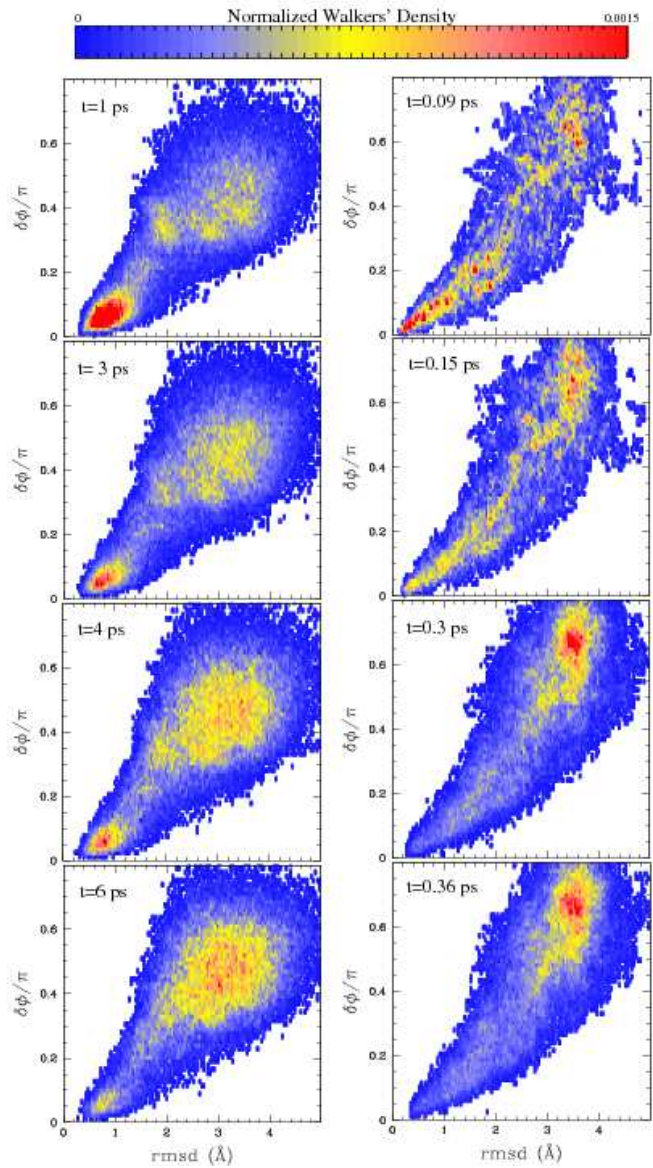}
\caption{\label{fig.t1} An effective migration of the walkers is observed on a time-scale 
$\tau\mathrm{fast} < t < \tau_\mathrm{slow}$, and it is signaled by the changes in the
relative density. Results shown here correspond to the walker density in different time
windows, for the initial condition A (left figures) and B (right figures), as defined in
Figure 2.}
\end{figure}

\begin{figure}
\includegraphics[width=\linewidth]{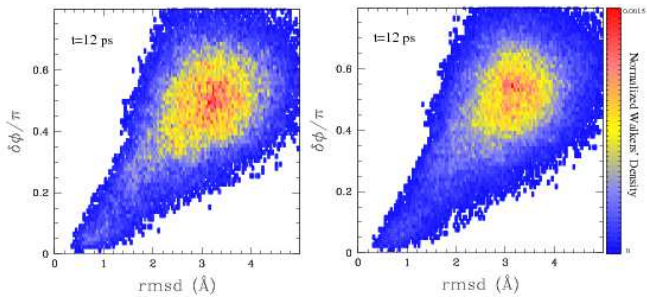}
\caption{\label{fig.t2} Over long time-scales $t\simeq \tau_\mathrm{slow}$ the
distribution of walkers reach equilibrium and the difference in the
initial conditions is completely lost. The left figure shows the walker density at
equilibrium for the initial condition A while the right figure corresponds to the initial 
condition B (see Fig.~\ref{fig.t0}).}
\end{figure}

The final result of our analysis is summarized in Fig.~\ref{fig.vecs}, where the selected
walkers are superimposed on the free energy profile independently determined by means of
extensive MD simulations and standard techniques. Remarkably, the walker
positions and the orientation of their compasses clearly highlight the minimum free
energy path connecting the native and the unfolded states. With a
more restrictive choice of the various threshold values, the transition state
region can be pinpointed as well. 
As predicted, the simulation time needed by the walkers to find the path is of
the order of $10^4$ time steps, 
significantly shorter than the characteristic time associated to activation
process, which is around $10^6$ time steps for the helix-coil
transition considered here. We
envision the time separation to be even more pronounced for more complex systems.

\section{Data analysis procedure}

\subsection{Effective migration and time window choice}

The reduced current we want to observe requires a time larger than
$\tau_\mathrm{fast}$ (although much smaller than $\tau_\mathrm{slow}$) to form.
On the other hand the current disappears once the equilibrium is achieved. 
A look to the evolution of the walker density helps fixing the most profitable
time window.  As an example, we show in Fig.~\ref{fig.t1} several snapshots of the walker
distribution obtained starting from the two different initial conditions 
displayed in Fig.~\ref{fig.t0}.
While case A reflects the Boltzmann distribution at $T_\mathrm{f}$, case B is
quite far from equilibrium. Figure~\ref{fig.t1} compares the time evolution of 
the walkers' density in the two cases. Finally, Fig.~\ref{fig.t2} shows that
when the equilibrium is reached any difference due to the different initial
conditions is lost.

Based upon the inspection of the walkers' migration, we select as time windows
the interval $[0.5,6]$ ps for case A, and $[0.09,0.36]$ ps for case B. The 
difference in the time scales of the walker's migration is due to the fact 
that initial condition B is at higher temperature than initial condition A.
 
\subsection{Density threshold and rate distribution}   

\begin{figure}
\includegraphics[width=\linewidth]{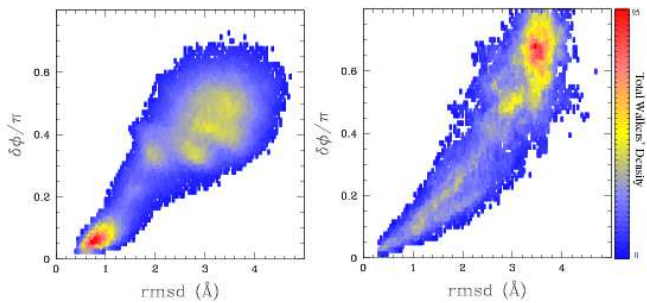}
\caption{\label{fig.dns} Walkers' density averaged over all the time snapshots
in the chosen time window, as discussed in the text. The left figure shows the walker 
density obtained for the initial condition A while the right figure corresponds to
the initial condition B (see Fig.~\ref{fig.t0}).}
\end{figure}

Once the time window is chosen, in order to reduce the unavoidable noise
present in the data, we filter out all the points of the grid that are not
consistently populated (i.e.~have a density below a given threshold) during the
migration process.  The distribution of the walkers' population in the space
spanned by the reaction coordinates shown in Fig.~\ref{fig.dns} is given by all the walker
configurations visited during all the independent simulations performed within
the considered time window.
In order to make the choice of the density threshold somewhat less
arbitrary, we adopt the following criterion: the threshold should be low enough
that we do not disconnect the two metastable states, but high enough to have a
fair statistics at each point of the grid. Within these two boundaries, we
verified that the actual value of the threshold does not affect the final result. 
After inspection of Fig.~\ref{fig.dns}, we choose the value 18 as density
threshold for case A and 16 for B. The selected configurations cover the native
state, the transition path and the unfolded state. Now we need some quantity to
discriminate between the states and the reaction path.

This is a good place to explain the mechanism of the walkers' effective
migration. If we picture the average rate $\mathcal{N}(\mathbf{w})$ for clonation/destruction
(defined in Eq.~(\ref{rate})), we notice that the probability of clonation is larger in
the unfolded state region (Fig.~\ref{fig.rate}). This drives the effective migration. One may notice
that the rate is everywhere negative: In fact, our implementation of the
supersymmetric Langevin equation is characterized by the fact that the number
of walkers grows exponentially. A random decimation of walkers when their
number exceeds some maximum value is sufficient to solve the problem and does
not introduce any significant bias in the final result.
 
\subsection{The reaction path revealed}

\begin{figure}
\includegraphics[width=\linewidth]{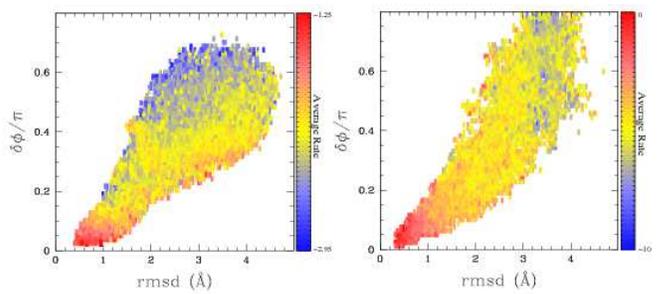}
\caption{\label{fig.rate} Distribution of the rate $\mathcal{N}(\mathbf{w})$ for
clonation or destruction of the walkers, averaged over all the snapshots, on the
2-dimensional space spanned by the reaction coordinates. Results shown in the 
left figure are obtained with the initial condition A while the right figure corresponds to 
the initial condition B (see Fig.~\ref{fig.t0}).}
\end{figure}

\begin{figure}
\includegraphics[width=\linewidth]{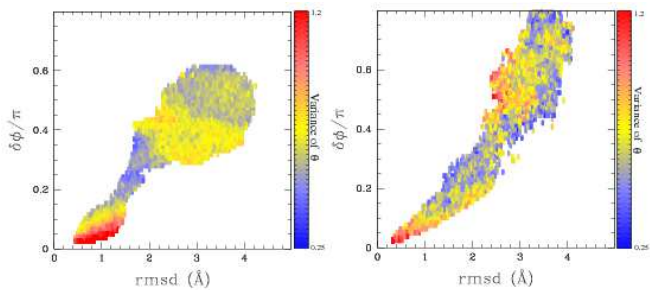}
\caption{\label{fig.varth} The average variance of the direction angle $\vartheta
= \arctan(\xi_2/\xi_1)$ (where $\xi_1,\xi_2$ are the reaction coordinates) indicates 
the region where the vectors associated to the walkers are more aligned. Results
shown in the left figure are obtained with the initial condition A while the right figure 
corresponds to the initial condition B (see Fig.~\ref{fig.t0}).}
\end{figure}

The walker density alone will not reveal the information we are most
interested in: that is, where the reduced current is stronger. This
information is stored in the ``compasses'' associated to the walkers: we 
expect the vectors to be strongly collinear in correspondence of the reaction
path, and disordered within the states. The averaged value of the vector is not
a reliable quantity to look at, because it can be affected by cancellations
between vectors with same direction but different sign. It is convenient
to define the direction angle $\vartheta\equiv\arctan(\xi_2/\xi_1)$, where
$\xi_1,\xi_2$ are the reaction coordinates we are using: Root mean square
deviation and helicity, respectively. The variance of $\vartheta$ is a good
measure of the coherence between the directions of vectors in the same cell of
our grid.  Figure~\ref{fig.varth} shows that the variance is indeed a good
marker for the reaction path. By combining the information of
Figures~\ref{fig.dns},~\ref{fig.rate},~\ref{fig.varth} we select a set of
walkers corresponding to densely populated regions, with an associated high
clonation rate, and with a small variance of $\vartheta$.

\begin{figure}
\includegraphics[width=\linewidth]{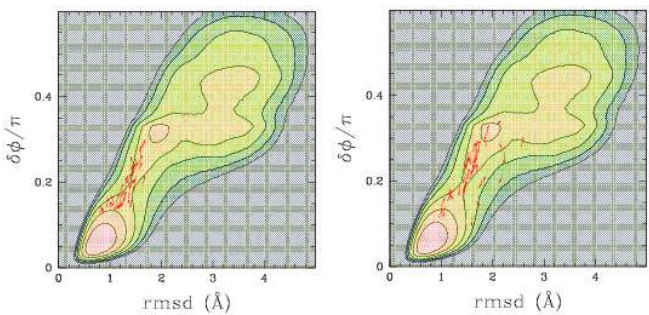}
\caption{\label{fig.vectors} The compasses of the selected walkers are 
compared with an independently derived free energy profile.
The left figure shows the results obtained for the initial condition A while the 
right figure corresponds to the initial condition B (see Fig.~\ref{fig.t0}). The 
red arrows illustrate the orientation of the compasses associated to the walkers,
and are superimposed to the independently determined free energy profile,
at the transition temperature $T_\mathrm{f}$.
Each contour marks an increase of free energy of 1 Kcal/mol.  Since
there are big differences in the length of these vectors (as discussed in the text)  
a logarithmic scale has been adopted to
improve the readability: If an arrow in the picture is twice longer, the actual norm 
of the vector is ten times larger.}
\end{figure}

Once a reasonable threshold is chosen for the variance of $\vartheta$, the
resulting configurations can be compared with the free energy profile
computed for the same model with traditional MD techniques. 
With a threshold value of 0.52 for case A and 0.45 for case B, the
result is shown in Fig.~\ref{fig.vectors}. 
Each vector displayed in the figure at a given position $(\xi_1, \xi_2)$
represents an average over a small volume $d\xi_1 d\xi_2$ centered in $(\xi_1, \xi_2)$.
All the figures were obtained with the values $d\xi_1 = 0.04$, and $d\xi_2 = 0.04$.

\section{Conclusions}

The results presented in this paper show that a supersymmetrically enhanced 
version of molecular dynamics can be efficiently used to identify transition 
states and reaction paths in models of macromolecular systems characterized 
by a clear separation of time-scales $\tau_\mathrm{slow} \gg \tau_\mathrm{fast}$. 
The great advantage of the method is that the simulation does not need to extend over the
long time-scale $\tau_\mathrm{slow}$, since the SuSy Kramers spectrum contains from
the very beginning all the information about the topology of the phase space
\cite{JSP122_557}. The trade-off is that instead of a single trajectory, a large number 
of walkers are used to explore the phase space. However, since each single
walker trajectory is extremely short, SuSy MD is easily and efficiently implemented in a parallel computing framework.

Althought the thoretical/mathematical foundation of the SuSy MD approach has some 
similarities with the recently proposed Finite Temperature String  (FTS) method
\cite{JPCB109_6688,JCP125_024106}, there are important differences. A 
comparison of these methods clearly highlights relative strengths and weaknesses.
While the FTS technique (as well as the transition path sampling \cite{ACP123_1}
that similarly relies on evolving a string rather than a point-like object in the
phase space) requires the definition of an initial and a final state, the SuSy walkers 
are able to find their own way without any previous knowledge of the configurational 
landscape. In addition, the SuSy approach does not require the FTS assumption 
that the isocommittor surface of the reaction could be locally approximated by
a hyperplane. On the other hand, FTS-based approaches bypass the problems 
related to the choice of reaction coordinates \cite{PNAS103_9885}, since they work 
directly in the high-dimensional phase space. Although the work presented in this
paper was based on the \emph{a priori} knowledge of a good set of reaction
coordinates, nothing in the method itself require such a step. It should be possible to 
modify the data analysis procedure in such a way that clusters of configurations along
the reaction path are read directly in the phase space. We believe this to be a promising direction for
further research.

\begin{acknowledgments}
We wish to thank Julien Tailleur for his kindness in sharing his experience,
and the NSF-funded Institute for Pure and Applied Mathematics at UCLA where
part of our work was performed.
This work has been supported in part by grants from NSF (C.C. Career CHE-0349303,
CCF-0523908, and CNS-0454333), and the
Robert A. Welch Foundation (C.C. Norman Hackerman Young Investigator award and
grant C-1570). The Rice Cray XD1 Cluster ADA used for the
calculations is funded by NSF under grant CNS-0421109, and a partnership between
Rice University, AMD, and Cray.
\end{acknowledgments}

\appendix*

\section{Explicit expression of the potential and values of the parameters}

The system selected for our study is a G\=o-like~\cite{JMB298_937} model that
uses a short alpha-helical segment as native structure. In particular, we chose
the first 12 residues of chain A of the Alanine-zipper described in
Ref.~\cite{JBC277_48708} (PDB code 1JCD) as native helical structure. 
The potential energy associated to the system is in the form:
\begin{widetext}
  \begin{eqnarray}
    V&=&k_\mathrm{r}\sum_{i=1}^{N_\mathrm{R}-1}(r_i-r^{(0)}_i)^2 +
    k_\theta\sum_{i=1}^{N_\mathrm{R}-2}(\theta_i-\theta_i^{(0)})^2 +
    \sum_{i=1}^{N_\mathrm{R}-3}\left[k_\varphi^{(1)}(1-\cos(\varphi_i-\varphi_i^{(0)})) +
    k_\varphi^{(2)}(1-\cos 3(\varphi_i-\varphi_i^{(0)}))\right] \nonumber \\
    & & 
    + \epsilon_1{\sum_{(i,j)\in C}}^{\prime}\left[ 5\left(\frac{\sigma_{ij}}{r_{ij}}\right)^{12} -
    6\left(\frac{\sigma_{ij}}{r_{ij}}\right)^{10}  \right]+\epsilon_2{\sum_{(i,j)\notin C}}^{\prime} 
    \left(\frac{\sigma_0}{r_{ij}}\right)^{12}  \,,
    \label{eq:pot}
  \end{eqnarray}
\end{widetext}
where:
\begin{itemize}
\item The residues are numbered from 1 to $N_\mathrm{R}$ and their
position is represented by the $C_\alpha$ atoms; 
\item $r_{ij}$ is
the distance between residues $i,j$, while $r_i\equiv r_{i,i+1}$;
\item $\theta_i$ is the angle between the vector from residue $i$ to $i+1$ and the vector from $i+1$ to $i+2$;
\item $\varphi_i$ is the dihedral angle formed by the residues $i,i+1,i+2,i+3$;
\item $\sum^{\prime}$ denotes a sum over pairs of residues $i,j$ with $j-i\ge 4$;
\item $C$ is the native contact map, that is the list of residue
pairs that are in contact in the native structure;
\item The constants $r_i^{(0)},\theta_i^{(0)},\varphi_i^{(0)}$ are
fixed by the corresponding values in the native structure;
\item The parameters $\sigma_{ij}$ are set equal to the distance
between the $C_{\alpha}$ atoms of residue $i$ and $j$ in the native structure.
\end{itemize}
The values of the remaining constants in Eq.(\ref{eq:pot}) have been chosen as follows:
\begin{center}
\begin{tabular}{| r @{=} c | r @{=} c | r @{=} c |}
\hline
$k_\mathrm{r}$ & 100 kcal/mol/\AA$^2$ & $k_{\varphi}^{(1)}$ & 1 kcal/mol   & $\epsilon_1$ & 5 kcal/mol  \\
\hline
$k_\theta$     &  20 kcal/mol/rad$^2$ & $k_{\varphi}^{(2)}$ & 0.5 kcal/mol & $\epsilon_2$ & 1 kcal/mol  \\
\hline
\end{tabular} 
\end{center}
while $\sigma_0$ is equal to 3 \AA, and the native contact map $C$ is
\begin{displaymath}
 \{(1,5),(2,6),(3,7),(4,8),(5,9),(6,10),(7,11),(8,12)\} \,.
\end{displaymath}
The transition temperature $T_\mathrm{f}$ of the system is defined as the temperature
corresponding to a peak in the heat capacity curve. With our choice of the
parameters we obtain $T_f \simeq 0.2 \epsilon_1/k_B$. All the thermodynamic
quantities (including the free energy surface reported in Fig.~\ref{fig.vecs}) were
obtained by combining extensive MD simulations at different temperature with
the Weighted Histogram Analysis Method (WHAM)~\cite{PRL61_2635,PRL63_1195}. 

In our implementation, the Langevin equation is solved by means of the second-order
quasi-symplectic integrator described in Ref.~\cite{PRE69_041107}, while the conservation 
of the norm of the vector $\mathbf{w}$ is achieved by applying the implicit 
midpoint rule (see, for instance, Ref.~\cite{Hairer2002}).

The friction coefficient entering the Langevin equation is set to $\gamma =
2.5$ ps$^{-1}$, and the mass of each particle is $m = 100$ amu.
The time step used in all dynamical simulations is $\delta t = 10^{-4}$ ps.
The SuSy MD simulations used $\sim 60,000$ independent walkers for each temporal
snapshot considered.

\bibliography{preversion}

\end{document}